\documentclass[intlimits,twoside,a4paper]{article}

\usepackage{amsmath,amssymb}
\usepackage{graphicx}
\usepackage{amsmath}

\usepackage[T2A]{fontenc}
\usepackage[cp1251]{inputenc}
\usepackage{cite}
\usepackage{epstopdf}

\usepackage[eqsecnum]{cmpj3}

\issue{2020}{23}{3}{33602}
\doinumber{10.5488/CMP.23.33602}
\title[Cahn-Hilliard model with Schl\"ogl reactions]%
{Cahn-Hilliard model with Schl\"ogl reactions: interplay of equilibrium and non-equilibrium phase
transitions. I. Travelling wave solutions}

\author[P.O. Mchedlov-Petrosyan,  L.N. Davydov]{P.O. Mchedlov-Petrosyan\footnote{\normalfont
peter.mchedlov@free.fr}, L.N. Davydov\footnote{\normalfont ldavydov@kipt.kharkov.ua}}

\address{A.I. Akhiezer Institute for Theoretical Physics,  National Science Center ``Kharkov Institite of
Physics \& Technology'',  1 Akademicheskaya Str., Kharkiv, Ukraine 61108  }
\date{Received March 17, 2020, in final form May 6, 2020}

\begin{document}

\maketitle

\begin{abstract}
The present work is devoted to the modelling which is based on the modified Cahn-Hilliard equation, the
interplay of equilibrium and non-equilibrium phase transitions. The non-equilibrium phase transitions
are modelled by the Schl\"ogl reactions systems. We consider the advancing fronts which combine these
both transitions. The traveling wave solutions are obtained; the conditions of their existence and
dependence on the parameters of the models are studied in detail. The possibility of the existance of
non-equilibrated phase  is discussed.

\end{abstract}

\section{Introduction}\label{Sec:1}

The present work is devoted to the modelling based on the  modified Cahn-Hilliard equation, the
interplay of equilibrium and non-equilibrium phase transitions. We consider the advancing fronts which
``combine'', in some sense, these both transitions. To understand the meaning of our modifications, we
need to give some insight into the history and into the existing modifications of this equation. The
Cahn-Hilliard equation \cite{1,2,3,4} is now a well-established model in the theory of phase
transitions as well as in several other fields. The basic underlying idea of this model is that for
inhomogeneous system, e.g., a system undergoing a phase transition, the thermodynamic potential (e.g., a
free energy) should depend not only on the order parameter $u$ but also on its gradient. The idea
of such dependence was already introduced  by Van der Waals \cite{5} in his theory of capillarity. For
an inhomogeneous system, the local chemical potential $\mu $ is defined as variational derivative of
 thermodynamic potential functional. If  thermodynamic potential is the simplest symmetric-quadratic-function of the gradient, this leads to the local chemical potential $\mu $ which depends on
Laplacian, while for the one-dimensional case it depends on the second order derivative of the order parameter.
The diffusional flux $J$ is proportional to the gradient of chemical potential $\nabla \mu $; the
coefficient of proportionality is called mobility $M$~\cite{6}. With such expression for the flux, the
diffusion equation instead of the usual second order equation becomes a forth-order PDE for the order
parameter $u$ (herein our notations differ from the notations in the original papers):
\begin{eqnarray} 
\frac{\partial u}{\partial t'} &=&\nabla \left[M\nabla \mu \right],\label{1.1} \\
\mu &=&-\bar{\varepsilon }^{2} \Delta u+f(u).
\label{ZEqnNum118824}
\end{eqnarray}
Here, $M$ is mobility, $\bar{\varepsilon }$ is usually assumed to be proportional to the capillarity
length, and $f(u)=\frac{\rd\Phi (u)}{\rd u} $, where $\Phi (u)$ is
homogeneous part of the thermodynamic potential. In the present communication, we  take
$f(u)$ in the form of the cubic polynomial (corresponding to the fourth-order polynomial
for the homogeneous part of thermodynamic potential):
\begin{equation} \label{ZEqnNum302882} f(u)=u^{3} -\delta u^{2} -su ;
\end{equation}
rescaling $u$, the coefficient at $u^{3} $ could be always scaled to one. In the present paper, $u$ is
assumed to be non-negative, so, even for a symmetric potential, $\delta \neq 0$; furthermore, the
asymmetric potential naturally appears in some modifications of the Cahn-Hilliard equation \cite{4}.
In this phenomenological model, we always consider  the isothermal situation, so we do not show the
temperature dependence of the coefficients in \eqref{ZEqnNum302882} explicitly. However, if we want to
model the approach to the critical state for such a model, the approach to critical temperature will be
manifested by merging the stationary states together, i.e., by two non-zero roots of the right-hand
side of \eqref{ZEqnNum302882} approaching the third zero root.

The classic Cahn-Hilliard equation was introduced as early as 1958 \cite{1,2}; the stationary
solutions were considered, the linearized version was treated and the corresponding instability of
homogeneous state was identified. However, an intensive study of the fully nonlinear form of this equation
started much later \cite{7}. At present, an impressive amount of work is done on nonlinear
Cahn-Hilliard equation\textbf{\textit{,}} as well as on its numerous modifications, see \cite{3,4}. An
important modification was done by Novick-Cohen \cite{8}. Taking into account the dissipation effects
which are neglected in the derivation of the classic Cahn-Hilliard equation, she introduced the
\textbf{\textit{viscous}} Cahn-Hilliard (VCH) equation
\begin{equation} \label{1.4} \frac{\partial u}{\partial t'} =\nabla \left[M\nabla \left(\mu +
\bar{\eta }\frac{\partial u}{\partial t'} \right)\right],
\end{equation}
where the coefficient $\bar{\eta }$ is called viscosity. It was also noticed that VCH equation could
be derived as a certain limit of the classic Phase-Field model \cite{9}. Later on, several authors
considered the nonlinear \textbf{\textit{convective}} Cahn-Hilliard equation (CCH) in one space
dimension \cite{10,11,12}
\begin{equation} \label{ZEqnNum429984} \frac{\partial u}{\partial t'} -\bar{\alpha }u\frac{\partial u}
{\partial x'} =\frac{\partial }{\partial x'} \left(\frac{\partial \mu }{\partial x'} \right).
\end{equation}
Leung \cite{10} proposed this equation as a continual description of lattice gas phase separation
under the action of an external field. Similarly, Emmott and Bray \cite{12} proposed this equation
as a model for the spinodal decomposition of a binary alloy in an external field \textit{E}. As they
noticed, if the mobility $M$~\cite{6} is independent of the order parameter (concentration), the term
involving \textit{E }will drop out of the dynamics. To get nontrivial results, they assumed the
simplest possible symmetric dependence of mobility on the order parameter, viz. $M\sim 1-ru^{2} $.
Then, they obtained the Burgers-type convection term in equation~\eqref{ZEqnNum429984} with the
coefficient $\bar{\alpha }=2rE$. Thus, the sign of $\bar{\alpha }$ depends both on the direction of
the field and on the sign of $r$. Witelski \cite{11} introduced the equation \eqref{ZEqnNum429984} as
a generalization of the classic Cahn-Hilliard equation or as a generalization of the
Kuramoto-Sivashinsky equation \cite{13,14} by including a nonlinear diffusion term. In
\cite{10,11,12}, and in \cite{15,16}, several approximate solutions and only two exact static kink and
anti-kink solutions were obtained. The ``coarsening'' of domains separated by kinks and by anti-kinks was
also discussed. To study the joint effects of nonlinear convection and viscosity, Witelski \cite{17}
introduced the convective-viscous-Cahn-Hilliard equation (CVCHE) with a general symmetric double-well
potential $\Phi (u)$:
\begin{equation} \label{ZEqnNum910197} \frac{\partial u}{\partial t'} -\bar{\alpha }u\frac{\partial u}
{\partial x'} =\frac{\partial }{\partial x'} \left[M\frac{\partial }{\partial x'} \left(\mu +\bar{\eta }
\frac{\partial u}{\partial t'} \right)\right],
\end{equation}
\begin{equation} \label{1.7} \mu =-\bar{\varepsilon }^{2} \frac{\partial ^{2} u}{\partial x^{'2} } +
\frac{\rd\Phi (u)}{\rd u}.
\end{equation}
It is worth noting that all results, including the stability of solutions, were obtained without
specifying a particular functional form of the potential. Thus, they are valid both for the polynomial
and logarithmic \cite{3,4} potential. Moreover, with a constraint imposed on nonlinearity and viscosity,
the approximate travelling-wave solutions were obtained. In \cite{18}, for equation
\eqref{ZEqnNum910197} with polynomial potential, see \eqref{ZEqnNum302882}, and the balance between
the applied field and viscosity, several exact single- and two-wave solutions were obtained.

Another modification of the nonlinear Cahn-Hilliard equation which attracted much interest is the
insertion of linear or nonlinear sink/source terms, e.g., due to a chemical reaction, into this
equation. Such a study was pioneered by Huberman \cite{19}. He introduced Cahn-Hilliard equation with
additional kinetic terms corresponding to the reversible first-order autocatalytic chemical reaction
and analyzed the linear stability of stationary states. Cohen and Murray \cite{20} considered the same
equation in the biological context: they used quadratic nonlinearity to describe the growth and dispersal
in the population model; they studied the stability and identified bifurcations to spatial structures.
Similar equation (with additional nonlinear term) was used in \cite{21} to study the segregation dynamics
of binary mixtures coupled with the chemical reaction. The same equation as in \cite{19,20} was used to
describe phase transitions in a chemisorbed layer \cite{22} and to model the system of cells that move,
proliferate and interact via adhesion \cite{23}. Furthermore, for the latter model, several rigorous
mathematical results on the existence and asymptotics of solutions were obtained \cite{24,25}. General
observation is that the presence of chemical reaction can visibly influence the equilibrium phase
transition, e.g., freeze the spinodal decomposition or coarsening, stabilizing some stationary
inhomogeneous state.

On the other hand, the canonical models for non-equilibrium phase transitions in chemical reaction
systems were introduced by Schl\"ogl \cite{26}; here, the different ``phases'' correspond to different
stationary states of the system. Schl\"ogl considered two reaction systems: the so-called ``First
Schl\"ogl Reaction''
\begin{equation} \label{ZEqnNum382394} A+X\rightleftarrows 2X,
\end{equation}
\begin{equation} \label{ZEqnNum219603} B+X\rightleftarrows C,
\end{equation}
and the ``Second Schl\"ogl Reaction''
\begin{equation} \label{ZEqnNum655012} A+2X\rightleftarrows 3X,
\end{equation}
\begin{equation} \label{ZEqnNum594450} B+X\rightleftarrows C.
\end{equation}
The concentrations of species $A$, $B$ and $C$ (which are called the ``reservoir reagents'') are
assumed to be constant and only concentration of $X$ can vary with time and space. For the first
Schl\"ogl reaction in the absence of diffusion, the evolution of $X$ is described by
\begin{equation} \label{ZEqnNum369199} \frac{\rd X}{\rd t} =\, -k'_{11} X^{2} +k_{11} AX-k_{21} BX+k'_{21} C.
\end{equation}
Here, the $k_{ij} \, ,\, \, k'_{ij} \, $ are the rate constants for the forward and reverse reactions,
respectively; the second lower index is ``1'' for the first Schl\"ogl reaction, and ``2'' for the
second one. Correspondingly, for the second Schl\"ogl reaction in the absence of diffusion, the
evolution of $X$ is described by
\begin{equation} \label{ZEqnNum324453} \frac{\rd X}{\rd t} =\, -k'_{12} X^{3} +k_{12} AX^{2} -k_{22} BX+k'_{22} C.
\end{equation}
The first reaction exhibits a non-equilibrium phase transition of the second order, the second reaction shows a phase
transition of the first order (for  details see \cite{26}). If the system simultaneously undergoes an
equilibrium phase transition accompanied by a phase separation, it could be of considerable interest to
study the interaction of an equilibrium and non-equilibrium phase transitions. Apparently, being  unaware of
Schl\"ogl paper, Huberman \cite{19} and Cohen and Murray \cite{20} in fact considered the interplay of
equilibrium and (the second-order) non-equilibrium phase transitions.

In the present communication, we consider the modified Cahn-Hilliard equation complemented by
source/sink terms corresponding both to the first and the second Schl\"ogl reactions. Let us call these
modifications Cahn-Hilliard-Huberman-Cohen-Murray (CHHCM) and Cahn-Hilliard-Schl\"ogl (CHS) equations,
respectively. We  also consider the influence of some additional modifications of the
Cahn-Hilliard equation, such as viscous and convective terms \cite{8,10,11,12,17,18}. We give exact
travelling-wave solutions for these modifications. For completeness in appendix we also give an exact
travelling-wave solution for Puri-Frish modification \cite{21}. In the second part of this paper, some
additional exact solutions and stability study are presented.

\section{Convective viscous Cahn-Hilliard-Huberman-Cohen-Murray equation}\label{Sec:2}

In the present section we first give exact travelling-wave solutions for convective viscous
Cahn-Hilliard equation with second order reaction terms. So, we first take into account the action
of both external field and dissipation \cite{8,10,11,12,17,18}; then, we drop the convective and
viscous terms, reducing equation to CHHCM equation. To avoid some unnecessary complications, we assume
reaction \eqref{ZEqnNum219603} to be irreversible, i.e., in \eqref{ZEqnNum369199} $k'_{21} =0$. In
terms of Schl\"ogl model \cite{26}, this corresponds to the ``analog of zero magnetic field'' case. From
\eqref{ZEqnNum910197}, \eqref{ZEqnNum118824}, \eqref{ZEqnNum302882} and \eqref{ZEqnNum369199} we write
down the Convective Viscous CHHCM equation, first in terms of the initial variable $X$
(concentration):
\begin{equation} \label{ZEqnNum718260}  {\frac{\partial X}{\partial t'} -
\bar{\alpha }X\frac{\partial X}{\partial x'} =M\frac{\partial ^{2} }{\partial x^{'2} } \left(\bar{\mu
}+\bar{\eta }\frac{\partial X}{\partial t'} \right)-} {k'_{11} X^{2} + k_{11} AX-k_{21} BX} ,
\end{equation}
\begin{equation} \label{2.2} \bar{\mu }=-\bar{\varepsilon }^{2} \frac{\partial ^{2} X}
{\partial x^{'2} } +\bar{f}\left(X\right),
\end{equation}
\begin{equation} \label{ZEqnNum307761} \bar{f}\left(X\right)=qX^{3} -\bar{\delta }X^{2} -\bar{s}X.
\end{equation}
The equations \eqref{ZEqnNum718260}--\eqref{ZEqnNum307761}  implicitly assume that in the system
$A-B-C-X$, the components $A$ and $B$ are in large excess, and they are not essentially exhausted during the
chemical reaction and did not  change essentially due to the phase transition; we  also assume $M$ to be
a constant. Renormalizing $\bar{X}$, $x'$ and $t'$, we introduce
\begin{equation} \label{2.4} X=uX_{0} ;\, \, \, x'=xl;\, \, \, t'=t\tau .
\end{equation}
Here, $X_{0} =\frac{1}{\sqrt{q} } $ ,$\, \tau =\frac{1}{k'_{11} X_{0} } =\frac{\sqrt{q} }{k'_{11} } $
and $l=\sqrt{M\tau } =\sqrt{\frac{M\sqrt{q} }{k'_{11} } } $. Denoting $\alpha =\bar{\alpha
}\frac{X_{0} \tau }{l} =\bar{\alpha }\sqrt{\frac{\sqrt{q} }{k'_{11} M} } $, $\varepsilon ^{2}
=\frac{\bar{\varepsilon }^{2} }{l^{2} } $, $\eta =\frac{\bar{\eta }}{\tau } $, $\delta
=\frac{\bar{\delta }}{X_{0} } =\, \bar{\delta }\sqrt{q} \, $ and $s=\bar{s}q$ we write down equation
\eqref{ZEqnNum718260} in the non-dimensional form,
\begin{equation} \label{ZEqnNum864329}  \frac{\partial u}{\partial t} -
\alpha u\frac{\partial u}{\partial x} =  {\frac{\partial ^{2} }{\partial x^{2} } \left(-\varepsilon
^{2} \frac{\partial ^{2} u}{\partial x^{2}}+u^{3} -\delta u^{2}-su+ \eta \frac{\partial u}{\partial
t} \right)-u\left(u-u_{1} \right)}.
\end{equation}
We  also introduce
\begin{equation} \label{ZEqnNum930329} u_{1} =\frac{k_{11} A-k_{21} B}{k'_{11} X_{0} } ,
\end{equation}
assuming $u_{1} >0$, i.e., $k_{11} A>k_{21} B$. Looking for the travelling wave solutions of
\eqref{ZEqnNum864329}, we introduce the travelling wave coordinate $z=x-vt$. This yields
 \begin{equation} \label{ZEqnNum308471} \frac{\rd}{\rd z} \left[vu+\alpha
\frac{u^{2} }{2} +\frac{\rd}{\rd z} \left(-\varepsilon ^{2} \frac{\rd^{2} u}{\rd z^{2} } +u^{3} - \delta u^{2}
-su-v\eta \frac{\rd u}{\rd z} \right)\right]= {u\left(u-u_{1} \right)} .
\end{equation}

We look for the solution, which connects the stationary state of the reaction system $u=u_{1} $ at
$z=-\infty $ with the stationary state $u=0$ at $z=+\infty $. The simplest possible ansatz
for the anti-kink solution (as usually we call ``kinks'' the solutions with $\frac{\rd u}{\rd z} >0$, and
``anti-kinks'' --- solutions with $\frac{\rd u}{\rd z} <0$) with this property is as follows:
\begin{equation} \label{ZEqnNum772972} \frac{\rd u}{\rd z} =\kappa u\left(u-u_{1} \right),
\end{equation}
where $\kappa $ is presently unknown positive constant. Then, equation \eqref{ZEqnNum308471} could be
written as
\begin{equation} \label{2.9} \frac{\rd}{\rd z} \left[vu+\alpha \frac{u^{2} }{2} -\frac{1}{\kappa }
u+\frac{\rd}{\rd z} \left(-\varepsilon ^{2} \frac{\rd^{2} u}{\rd z^{2} } +u^{3} -\delta u^{2} -su-v\eta
\frac{\rd u}{\rd z} \right)\right]=0.
\end{equation}
Integrating once, we get
\begin{equation} \label{2.10} vu+\alpha \frac{u^{2} }{2} -\frac{1}{\kappa } u+\frac{\rd}{\rd z}
\left(-\varepsilon ^{2} \frac{\rd^{2} u}{\rd z^{2} } +u^{3} -\delta u^{2} -su-v\eta \frac{\rd u}{\rd z}
\right)=C_{1} .
\end{equation}
Regarding the ansatz \eqref{ZEqnNum772972}, for the latter equation to be satisfied the expression
under the derivative should be proportional to $u$. That is, for \eqref{ZEqnNum772972} to give the
solution of \eqref{ZEqnNum864329}, two equations should be satisfied for arbitrary $u$
\begin{equation} \label{ZEqnNum831739} vu+\alpha \frac{u^{2} }{2} -\frac{1}{\kappa } u+\beta
\frac{\rd u}{\rd z} =C_{1} ,
\end{equation}
\begin{equation} \label{ZEqnNum913589} -\varepsilon ^{2} \frac{\rd^{2} u}{\rd z^{2} } +u^{3} -
\delta u^{2} -su-v\eta \frac{\rd u}{\rd z} =\beta u+C_{2}\, ,
\end{equation}
where $\beta ,\, \, C_{1} $ and $C_{2} $ are constants. The expression for the second derivative of
$u$ is easily written as:
\begin{equation} \label{ZEqnNum572871} \; \frac{\rd^{2} u}{\rd z^{2} } =\kappa ^{2}
\left(2u^{3} -3u_{1} u^{2} +u_{1}^{2} u\right).
\end{equation}
Then, equations~\eqref{ZEqnNum831739}, \eqref{ZEqnNum913589} take the form
\begin{equation} \label{ZEqnNum978156} \left(\frac{\alpha }{2} +\beta \kappa \right)u^{2} +
\left(v-\frac{1}{\kappa } -\beta \kappa u_{1} \right)u=C_{1} ,
\end{equation}
\begin{equation} \label{ZEqnNum927771} {-\varepsilon ^{2} \kappa ^{2}
\left[2u^{3} -3u_{1} u^{2} +u_{1}^{2} u\right]+} {u^{3} -\delta u^{2} -\left(s+\beta \right)u-v\eta
\kappa \left(u^{2} -u_{1} u\right)=C_{2} } \;  .
\end{equation}

Rearranging the terms and equating the coefficients at each power of $u$ to zero, we finally obtain five
constraints on the parameters:
\begin{equation} \label{ZEqnNum395401} \frac{\alpha }{2} +\beta \kappa =0 ,
\end{equation}
\begin{equation} \label{ZEqnNum533099} v=\frac{1}{\kappa } +\beta \kappa u_{1} ,
\end{equation}
\begin{equation} \label{ZEqnNum118687} \kappa ^{2} =\frac{1}{2\varepsilon ^{2} } ,
\end{equation}
\begin{equation} \label{2.19} v\eta \kappa =\frac{3}{2} u_{1} -\delta ,
\end{equation}
\begin{equation} \label{ZEqnNum266062} v\eta \kappa u_{1} =\frac{1}{2} u_{1}^{2} +s+\beta .
\end{equation}

There are five constraints \eqref{ZEqnNum395401}--\eqref{ZEqnNum266062} and only three unknowns $\kappa
,\, \, v$ and $\beta $. That is, for the constant velocity transition front to exist, two additional
constraints on the values of the stationary states of the reaction system and on the values of the
equilibrium states for the phase transition should be imposed. Now, there is some freedom in selecting
which parameters are ``basic'', those related to the reaction system, or those related to the
``Cahn-Hilliard part''. Assuming the former to be basic, we write the constraints as
\begin{equation} \label{ZEqnNum977184} \delta =\frac{u_{1} }{2} \left(3+\frac{\alpha \eta }
{\sqrt{2} \varepsilon } \right)-\eta ,
\end{equation}
 \begin{equation} \label{ZEqnNum223277} s=-\frac{u_{1}^{2} }{2} \left(1+\frac{\alpha \eta }{\sqrt{2} \varepsilon } \right)+\eta u_{1}
+\frac{\alpha \varepsilon }{\sqrt{2} } \, .
\end{equation}

If the constraints \eqref{ZEqnNum395401}--\eqref{ZEqnNum266062} are satisfied, the solution of equation
\eqref{ZEqnNum772972} is simultaneously the solution of the travelling-wave equation
\eqref{ZEqnNum308471}. Integrating \eqref{ZEqnNum772972} once, we get
\begin{equation} \label{ZEqnNum497203} u=\frac{u_{1} \exp \left\{-\kappa u_{1} \left(z+\phi
\right)\right\}}{1+\exp \left\{-\kappa u_{1} \left(z+\phi \right)\right\}} ,
\end{equation}
where $\phi $ is an arbitrary constant. It is natural to take position of the maximal value of the
derivative $\frac{\rd u}{\rd z} $ (when $\frac{\rd^{2} u}{\rd z^{2} } =0$), as $z=0$; then, $\phi =0$. The solution
\eqref{ZEqnNum497203} could be rewritten in the form
\begin{equation} \label{ZEqnNum151330} u=\frac{u_{1} }{2} \left[1-\tanh\left(\frac{u_{1} }{2\sqrt{2}
\varepsilon } \left(x-vt\right)\right)\right].
\end{equation}
Here, we used $\kappa =\frac{1}{\sqrt{2} \varepsilon } $, see \eqref{ZEqnNum118687}; the velocity
$v$ of  the transition front is given by \eqref{ZEqnNum395401} and \eqref{ZEqnNum533099},
 \begin{equation} \label{ZEqnNum363435} v=\sqrt{2} \varepsilon +\beta \kappa u_{1} =\sqrt{2}
\varepsilon -\frac{1}{2} \alpha u_{1} .
\end{equation}

The roots of equation
\begin{equation} \label{2.26} \tilde{u}\left(\tilde{u}^{2} -\delta \tilde{u}-s\right)=0
\end{equation}
correspond to the extrema of the homogeneous part of  thermodynamic potential
\eqref{ZEqnNum118824}, \eqref{ZEqnNum302882}, $\tilde{u}_{1} ,\, \, \tilde{u}_{3} $ are stable minima
and $\tilde{u}_{2} $ is unstable maximum. The root $\tilde{u}_{3} =0$ coincides with one of the
stationary states of the reaction system. Substitution of \eqref{ZEqnNum977184} and
\eqref{ZEqnNum223277} into the latter equation for $\delta $ and $s$, respectively, yields two
remaining roots, i.e., two constraints imposed on the values of $\tilde{u}_{1} ,\, \, \tilde{u}_{2} $
and $u_{1} $,
\begin{equation} \label{ZEqnNum844082} \tilde{u}_{1,2} =\frac{1}{2} \left[\frac{u_{1} }{2}
\left(3+\frac{\alpha \eta }{\sqrt{2} \varepsilon } \right)-\eta \pm \sqrt{G} \right],
\end{equation}
\begin{equation} \label{ZEqnNum353049} G=\left[\frac{u_{1} }{2} \left(3+\frac{\alpha \eta }{\sqrt{2}
\varepsilon } \right)-\eta \right]^{2} +4\left[-\frac{u_{1}^{2} }{2} \left(1+\frac{\alpha
\eta }{\sqrt{2} \varepsilon } \right)+\eta u_{1} +\frac{\alpha \varepsilon }{\sqrt{2} } \right].
\end{equation}
Here, the discriminator of quadratic equation is denoted by $G$ for convenience. To understand the
mutual effect of the equilibrium and non-equilibrium transitions, it is practical to consider
several special cases of \eqref{ZEqnNum844082}--\eqref{ZEqnNum353049}. First we consider the CHHCM
case, i.e., the absence of the applied field and dissipation. For $\alpha =0$ and $\eta =0$, expression
\eqref{ZEqnNum353049} simplifies drastically, yielding $G=\frac{1}{4} u_{1}^{2} $. Then,
\eqref{ZEqnNum844082} becomes
\begin{equation} \label{2.29} \tilde{u}_{1,2} =\frac{u_{1} }{4} \left(3\pm 1\right).
\end{equation}

This means that for the constant-velocity-transition front to exist, the values of the order
parameter corresponding to the equilibrium phases should coincide exactly with the values
corresponding to the stationary states of the chemical reactions system, i.e., $\tilde{u}_{1} =u_{1}
;\, \, \, \tilde{u}_{3} =0$. The thermodynamic potential should be symmetric, $\tilde{u}_{2}
={\tilde{u}_{1} \mathord{\left/ {\vphantom {\tilde{u}_{1}  2}} \right. \kern-\nulldelimiterspace} 2}
$, with equal-depth wells. The velocity depends on the $\varepsilon $ only, $v=\sqrt{2} \varepsilon $.
Now, let $\alpha =0;\, \, \, \eta \ne 0$. From \eqref{ZEqnNum353049} it follows
\begin{equation} \label{2.30} G=\left(\frac{u_{1} }{2} +\eta \right)^{2} ;\, \, \,
\tilde{u}_{1,2} =\frac{1}{2} \left[\frac{3u_{1} }{2} -\eta \pm \left(\frac{u_{1} }{2} +\eta \right)\right].
\end{equation}
That is, stationary values for the equilibrium transition should again coincide with the
stationary values for the reaction system, but the unstable value should be shifted to the lower value.
As it was mentioned in the introduction, the derivative of the homogeneous part of  thermodynamic
potential $\Phi (u)$ is given by~\eqref{ZEqnNum302882}:
\begin{equation} \label{2.31} \frac{\rd\Phi (u)}{\rd u} =u^{3} -\delta u^{2} -su.
\end{equation}

Integrating once and substituting values of $\delta $ and $s$ for $\alpha =0$, we obtain the following
expressions for the potential values $\Phi \left(\tilde{u}_{1} \right)$ and $\Phi \left(\tilde{u}_{3}
\right)$
\begin{equation} \label{ZEqnNum545762} \Phi \left(\tilde{u}_{1} \right)=-\, \frac{\eta }{6} u_{1}^{3}
\, +C;\, \, \, \Phi \left(\tilde{u}_{3} \right)=\Phi \left(0\right)=C.
\end{equation}
That is, to compensate the dissipation, the potential well corresponding to $\tilde{u}_{1} $ should be
deeper. On the other hand, if $\, \alpha \ne 0;\, \, \, \eta =0$
\begin{equation} \label{ZEqnNum472305} G=\frac{u_{1}^{2} }{4} +4\frac{\alpha \varepsilon }{\sqrt{2} } ;
\, \, \, \tilde{u}_{1,2} =\frac{u_{1} }{4} \left[3\pm \sqrt{1+16\frac{\alpha \varepsilon }
{\sqrt{2} u_{1}^{2} } } \right].
\end{equation}
This means that for positive $\alpha $, the order parameter value for the final state after
transition, $u=u_{1} $, is somewhat lower than the equilibrium value $\tilde{u}_{1} $. To ensure the
positivity of $\tilde{u}_{2} $ it should be $\sqrt{2} \alpha \varepsilon <u_{1}^{2} $; however, the
parameter $\varepsilon $ is small, so it is not a severe limitation. Now, let both $\alpha \ne 0$,
$\eta \ne 0$. The expression for the velocity \eqref{ZEqnNum363435} is independent of $\eta $; for the
special value $\alpha =\frac{2\sqrt{2} }{u_{1} } \varepsilon $, the velocity is zero, i.e., for the
corresponding value of the applied field, the transition front becomes static. Substitution of this
value of $\alpha $ into \eqref{ZEqnNum844082} and \eqref{ZEqnNum353049} yields
\begin{equation} \label{ZEqnNum160622} \tilde{u}_{1,2} =\frac{u_{1} }{4} \left[3\pm
\sqrt{1+\frac{32\varepsilon ^{2} }{u_{1}^{3} } } \right].
\end{equation}

Interestingly, the viscosity $\eta $ has dropped out from the latter expression. This is physically
reasonable: there is no dissipation for the static transition front; the deviation of the order
parameter value $u=u_{1} $ for the final state after transition from its equilibrium value
$\tilde{u}_{1} $ is exactly the same as given by \eqref{ZEqnNum472305} (i.e., for $\eta =0$ case) for
this special value of $\alpha $.

\section{Convective viscous Cahn-Hilliard-Schl\"ogl equation}\label{Sec:3}

In this section we first give exact travelling-wave solutions for a convective viscous Cahn-Hilliard
equation with third order reaction terms. Again, we first take into account the effect of both
external field and dissipation \cite{8,10,11,12,17,18}; then, we drop the convective and viscous terms,
reducing the equation to CHS equation. To make the calculations somewhat more transparent we assume the
reaction \eqref{ZEqnNum594450} to be irreversible, i.e., in \eqref{ZEqnNum324453} $k'_{22} =0$. From
\eqref{ZEqnNum910197}, \eqref{ZEqnNum118824}, \eqref{ZEqnNum302882} and \eqref{ZEqnNum324453}, we write
down the Convective Viscous CHS equation, first in terms of the initial variable $X$ (concentration):
\begin{equation} \label{ZEqnNum504256} {\frac{\partial X}{\partial t'} -
\bar{\alpha }X\frac{\partial X}{\partial x'} =M\frac{\partial ^{2} }{\partial x^{'2} } \left(\bar{\mu
}+\bar{\eta }\frac{\partial X}{\partial t'} \right)-} {k'_{12} X^{3} +k_{12} AX^{2} -k_{22} BX} \, ,
\end{equation}
\begin{equation} \label{3.2} \bar{\mu }=-\bar{\varepsilon }^{2} \frac{\partial ^{2} X}{\partial x^{'2} } +
\bar{f}\left(X\right),
\end{equation}
\begin{equation} \label{ZEqnNum518642} \bar{f}\left(X\right)=qX^{3} -\bar{\delta }X^{2} -\bar{s}X.
\end{equation}
Writing down equations \eqref{ZEqnNum504256}--\eqref{ZEqnNum518642}, we again assume implicitly that in
the system $A-B-C-X$ the components $A$ and $B$ are in large excess and are not essentially exhausted
during the chemical reaction; we  also assume $M$ to be a constant. Renormalizing $X$, $x'$ and
$t'$, we introduce
\begin{equation} \label{3.4} X=uX_{0} ;\, \, \, x'=xl;\, \, \, t'=t\tau .
\end{equation}
Here, $X_{0} =\frac{1}{\sqrt{q} } $, $\tau =\frac{1}{k'_{12} X_{0}^{2} } =\frac{q}{k'_{12} } $ and
$l=\sqrt{M\tau } =\sqrt{\frac{M}{k'_{12} X_{0}^{2} } } =\sqrt{\frac{Mq}{k'_{12} } } $. Denoting
$\alpha =\bar{\alpha }\frac{X_{0} \tau }{l} =\bar{\alpha }\frac{1}{\sqrt{k'_{12} M} }  $;
$\varepsilon ^{2} =\frac{\bar{\varepsilon }^{2} }{l^{2} } $; $\eta =\frac{\bar{\eta }}{\tau } $;
$\delta =\frac{\bar{\delta }}{X_{0} } =\, \bar{\delta }\sqrt{q} \, $; $s=\bar{s}q$; $R=\frac{k_{12}
A}{k'_{12} X_{0} } $ and $Q=\frac{k_{22} B}{k'_{12} X_{0}^{2} } $, we write down equation
\eqref{ZEqnNum504256} in non-dimensional form
\begin{equation} \label{ZEqnNum688383} {\frac{\partial u}{\partial t} -\alpha u
\frac{\partial u}{\partial x} =\frac{\partial ^{2} }{\partial x^{2} } \left(-\varepsilon ^{2}
\frac{\partial ^{2} u}{\partial x^{2} } +u^{3} -\delta u^{2} -su+\eta \frac{\partial u} {\partial t}
\right)-} {u\left(u^{2} -Ru+Q\right)} .
\end{equation}

Herein below we assume that the quadratic equation
\begin{equation} \label{3.6} u^{2} -Ru+Q=0
\end{equation}
always has real roots $u_{1} ,\, u_{2} ,\, \, u_{1} \geqslant u_{2} $; i.e. $R^{2} -4Q\geqslant 0$ which means, in
terms of the parameters of the reaction system, $\left(k_{12} A\right)^{2} \geqslant 4k'_{12} k_{22} B$.
Looking for the travelling wave solutions of \eqref{ZEqnNum688383}, we introduce the travelling wave
coordinate $z=x-vt$. This yields
\begin{equation} \label{ZEqnNum237678}  \frac{\rd}{\rd z} \left[vu+\alpha \frac{u^{2} }{2} +
\frac{\rd}{\rd z} \left(-\varepsilon ^{2} \frac{\rd^{2} u}{\rd z^{2} } +u^{3} -\delta u^{2} -su-v\eta
\frac{\rd u}{\rd z} \right)\right]= {u\left(u-u_{1} \right)\left(u-u_{2} \right)} \, .
\end{equation}

As in the previous section, we look for the solution, which connects the stationary state of the
reaction system $u=u_{1} $ at $z=-\infty $ with the stationary state $u=0$ at $z=+\infty $. Thus, the
proper ansatz for the anti-kink solution is again \eqref{ZEqnNum772972}
\begin{equation} \label{ZEqnNum442797} \frac{1}{\kappa } \frac{\rd u}{\rd z} =u\left(u-u_{1} \right),
\end{equation}
where $\kappa $ is presently an unknown positive constant. Then, equation \eqref{ZEqnNum237678} could be
written as
\begin{equation} \label{3.9} \frac{\rd}{\rd z} \left[vu+\alpha \frac{u^{2} }{2} +\frac{\rd}{\rd z}
\left(-\varepsilon ^{2} \frac{\rd^{2} u}{\rd z^{2} } +u^{3} -\delta u^{2} -su-v\eta \frac{\rd u}{\rd z}
\right)\right] =\frac{\rd}{\rd z} \left(\frac{1}{2\kappa } u^{2} -\frac{u_{2} }{\kappa } u\right) .
\end{equation}
Integrating once, we get
\begin{equation} \label{3.10} \left(v+\frac{u_{2} }{\kappa } \right)u+\left(\alpha -\frac{1}{\kappa }
\right)\frac{u^{2} }{2} +\frac{\rd}{\rd z} \left(-\varepsilon ^{2} \frac{\rd^{2} u}{\rd z^{2} } +u^{3} -
\delta u^{2} -su-v\eta \frac{\rd u}{\rd z} \right)=C_{1} .
\end{equation}

Regarding the ansatz \eqref{ZEqnNum442797}, for the latter equation to be satisfied, the expression
under the derivative should be proportional to $u$. That is, for \eqref{ZEqnNum442797} to give the
solution of \eqref{ZEqnNum237678} two equations should be satisfied for arbitrary $u$,
\begin{equation} \label{ZEqnNum737857} \left(v+\frac{u_{2} }{\kappa } \right)u+\left(\alpha -
\frac{1}{\kappa } \right)\frac{u^{2} }{2} +\beta \frac{\rd u}{\rd z} =C_{1}\,,
\end{equation}
\begin{equation} \label{ZEqnNum662757} -\varepsilon ^{2} \frac{\rd^{2} u}{\rd z^{2} } +u^{3} -
\delta u^{2} -su-v\eta \frac{\rd u}{\rd z} =\beta u+C_{2}\, ,
\end{equation}
where $C_{1} ,\, \, C_{2} $ and $\beta $ are constants. If the above constraints are satisfied for
arbitrary $u$, the solution of~\eqref{ZEqnNum688383} is again given by \eqref{ZEqnNum151330}, though
with different values of $u_{1} ,\, \, v,\, \, \varepsilon _{}^{} $. The expression for the second
derivative of $u$ is given again by \eqref{ZEqnNum572871}. Then, equations \eqref{ZEqnNum737857},
\eqref{ZEqnNum662757} take the form
\begin{equation} \label{3.13} \left(v+\frac{u_{2} }{\kappa } \right)u+\left(\alpha -\frac{1}{\kappa }
\right)\frac{u^{2} }{2} +\beta \kappa \left(u^{2} -u_{1} u\right)=C_{1} \,,
\end{equation}
\begin{equation} \label{3.14} {-\varepsilon ^{2} \kappa ^{2} \left(2u^{3} -3u_{1}
u^{2} +u_{1}^{2} u\right)+u^{3} -\delta u^{2}} {-\left(s+\beta \right)u-v\eta \kappa \left(u^{2}
-u_{1} u\right)=C_{2} } \, .
\end{equation}

Rearranging the terms and equating coefficients at each power of $u$ to zero, we finally obtain five
constraints on the parameters:
\begin{equation} \label{ZEqnNum207470} \kappa ^{2} =\frac{1}{2\varepsilon ^{2} } \, \, ,
\end{equation}
\begin{equation} \label{3.16} \frac{3}{2} u_{1} -\delta -v\eta \kappa =0,
\end{equation}
\begin{equation} \label{3.17} -\frac{1}{2} u_{1}^{2} -s-\beta +v\eta \kappa u_{1} =0,
\end{equation}
\begin{equation} \label{3.18} \beta =\frac{1}{2\kappa ^{2} } -\frac{\alpha }{2\kappa } ,
\end{equation}
\begin{equation} \label{ZEqnNum428799} v=\, \beta \kappa u_{1} -\frac{u_{2} }{\kappa }
\end{equation}

Similarly to \eqref{ZEqnNum395401}--\eqref{ZEqnNum266062}, there are five constraints
\eqref{ZEqnNum207470}--\eqref{ZEqnNum428799} and only three unknowns $\kappa ,\, \, v$ and $\beta $.
That is, for the constant velocity transition front to exist, two additional constraints on the values of
the stationary states of the reaction system and on the values of the equilibrium states for the phase
transition should be imposed. Assuming, as in section~\ref{Sec:2}, the parameters related to reaction
system to be ``basic'', we write the constraints as
\begin{equation} \label{ZEqnNum719941} \delta =\frac{3}{2} u_{1} -\left(\frac{u_{1} }{2} -u_{2} \right)\eta +
\frac{\alpha u_{1} }{2\sqrt{2} \varepsilon } \eta ;
\end{equation}
\begin{equation} \label{ZEqnNum552588} s=-\frac{1}{2} u_{1}^{2} -\varepsilon ^{2} +
\left(\frac{u_{1}^{2} }{2} -u_{1} u_{2} \right)\eta +\frac{\alpha }{\sqrt{2} } \left(\,
\varepsilon -\frac{u_{1}^{2} }{2\varepsilon } \eta \right).
\end{equation}
The latter expressions impose evident limitations on the roots of
\begin{equation} \label{3.22} \tilde{u}\left(\tilde{u}^{2} -\delta \tilde{u}-s\right)=0,
\end{equation}
i.e., on the extrema of the homogeneous part of the thermodynamic potential \eqref{ZEqnNum118824},
\eqref{ZEqnNum302882}, here, $\tilde{u}_{1} ,\, \, \tilde{u}_{3} $ correspond to stable minima and
$\tilde{u}_{2} $ to unstable maximum. The root $\tilde{u}_{3} =0$ coincides with one of the stationary
states of the reaction system. The expressions for two remaining roots yield two constraints imposed
on the values of $\tilde{u}_{1} ,\, \, \tilde{u}_{2} $ and $u_{1} $. The velocity of the transition
front $v$ is
\begin{equation} \label{ZEqnNum386863} v=\sqrt{2} \varepsilon \left(\frac{u_{1} }{2} -u_{2}
\right)-\frac{\alpha u_{1} }{2}
\end{equation}

To understand the mutual effect of the equilibrium and non-equilibrium transitions it is again
practical to consider several special cases of \eqref{ZEqnNum719941} and \eqref{ZEqnNum552588}. First
we consider the CHS case, i.e., the absence of the applied field and dissipation. For $\alpha =0$ and
$\eta =0$, these expressions simplify drastically, yielding
\begin{equation} \label{ZEqnNum335639} \delta =\frac{3}{2} u_{1} ;\, \, \, s=-\frac{1}{2} u_{1}^{2} -
\varepsilon ^{2}
\end{equation}
and, correspondingly
\begin{equation} \label{ZEqnNum238118} \tilde{u}_{1,2} =\frac{u_{1} }{4} \left(3\pm \sqrt{1-
\frac{16\varepsilon ^{2} }{u_{1}^{2} } } \right)\simeq \frac{u_{1} }{4} \left[3\pm \left(1-8
\frac{\varepsilon ^{2} }{u_{1}^{2} } \right)\right].
\end{equation}
That is, even in the absence of the applied field and viscosity, the order parameter value for the final
state after transition, $u=u_{1} $, is somewhat higher than the equilibrium value $\tilde{u}_{1} $.
The velocity is
\begin{equation} \label{ZEqnNum959690} v=\sqrt{2} \varepsilon \left(\frac{u_{1} }{2} -u_{2} \right).
\end{equation}

Remarkably, the dependence of velocity on the stationary values of concentration, $u_{1} ,\, \, u_{2}
,\, \, 0,$ is exactly the same as for the well known travelling-wave solution for the diffusion
equation with cubic nonlinearity; for $u_{2} ={u_{1} }/{2} $, the velocity is zero, that is the
front becomes static. However, the coefficient in \eqref{ZEqnNum959690} depends on $\varepsilon $,
i.e., on the ``Cahn-Hilliard part''. As it was mentioned in the introduction, the derivative of the
homogeneous part of the thermodynamic potential $\Phi (u)$ is given by
\eqref{ZEqnNum302882}:
\begin{equation} \label{3.27} \frac{\rd\Phi (u)}{\rd u} =u^{3} -\delta u^{2} -su.
\end{equation}
Integrating once and substituting values of $\delta $ and $s$ given by \eqref{ZEqnNum335639}, we obtain
the following expression for the potential $\Phi (u)$
\begin{equation} \label{ZEqnNum155679} \Phi (u)=\frac{1}{4} u^{4} -\frac{1}{2} u_{1} u^{3} +
\frac{1}{2} \left(\frac{1}{2} u_{1}^{2} +\varepsilon ^{2} \right)u^{2} +C,
\end{equation}
where $C$ is a constant. Then, final (after transition) value of the potential is $\Phi \left(u_{1}
\right)={1}/{2} \varepsilon ^{2} u_{1}^{2} +C$. Taking into account $\varepsilon \ll 1$, to
calculate the equilibrium value $\Phi \left(\tilde{u}_{1} \right)$, we use the approximate expression
from \eqref{ZEqnNum238118}, $\tilde{u}_{1} \simeq u_{1} -2{\varepsilon ^{2} }/{u_{1} } $.
Substitution into \eqref{ZEqnNum155679} and neglecting higher order in $\varepsilon ^{2} $ terms,
yields $\Phi \left(\tilde{u}_{1} \right)\simeq {1}/{2} \varepsilon ^{2} u_{1}^{2} +C$, i.e., it is
nearly equal to the value after transition.

It means that despite the deviation of the concentration in the final state after transition from its
equilibrium value, the deviations of thermodynamic potential from its equilibrium value are of the
higher order in $\varepsilon ^{2} $. Now, let $\alpha =0;\, \, \, \eta \ne 0$ in \eqref{ZEqnNum719941},
\eqref{ZEqnNum552588},
\begin{equation} \label{ZEqnNum316998} \tilde{u}_{1,2} =\frac{1}{2} \left\{\frac{3}{2} u_{1} -
\left(\frac{u_{1} }{2} -u_{2} \right)\eta \pm \sqrt{\left[\frac{u_{1} }{2} +\left(\frac{u_{1} }{2} -u_{2}
\right)\eta \right]^{2} -4\varepsilon ^{2} } \right\}.
\end{equation}

From \eqref{ZEqnNum959690} $\left(\frac{u_{1} }{2} -u_{2} \right)=\frac{v}{\sqrt{2} \varepsilon } $;
comparing \eqref{ZEqnNum238118} and \eqref{ZEqnNum316998} we see that the deviation term is of the form
$\frac{v\eta }{\sqrt{2} \varepsilon } $, i.e., multiple of velocity and viscosity. On the other hand,
if $\, \alpha \ne 0;\, \, \, \eta =0$
\begin{equation} \label{ZEqnNum106708} \tilde{u}_{1,2} =\frac{u_{1} }{4} \left[3\pm \sqrt{1+
\frac{16\varepsilon }{u_{1}^{2} } \left(\frac{\alpha }{\sqrt{2} } -\varepsilon \right)} \right].
\end{equation}

Now, let both $\alpha $ and $\eta $  be non-zero. The expression for the velocity
\eqref{ZEqnNum386863} is independent of $\eta $; for the special value of $\alpha $,
\begin{equation} \label{ZEqnNum936193} \alpha =\frac{2\sqrt{2} \varepsilon }{u_{1} }
\left(\frac{u_{1} }{2} -u_{2} \right),
\end{equation}
the velocity is zero, i.e., for the corresponding value of the applied field, the transition front
becomes static even for $u_{2} \ne \frac{u_{1} }{2} $. Substitution of this value of $\alpha $ into
\eqref{ZEqnNum719941} and \eqref{ZEqnNum552588} yields
\begin{equation} \label{3.32} s=-\frac{1}{2} u_{1}^{2} -\frac{2\varepsilon ^{2} u_{2} }{u_{1} } ;
\quad \delta =\, \frac{3}{2} u_{1} \, ,
\end{equation}
and
\begin{equation} \label{ZEqnNum383371} \tilde{u}_{1,2} =\frac{1}{4} u_{1} \left(3\pm \sqrt{1-32
\frac{\varepsilon ^{2} u_{2} }{u_{1}^{3} } } \right).
\end{equation}

Again, the viscosity $\eta $ has self-consistently dropped out of the latter expression, there is no
dissipation for the static transition front; the deviation of the order parameter value $u_{1} $ after
transition from its equilibrium value $\tilde{u}_{1} $ is exactly the same as given by
\eqref{ZEqnNum106708} (i.e., for $\eta =0$ case) for this special value of~$\alpha $.

\section{Discussion}\label{Sec:4}

In the present work we have modelled the interplay of equilibrium and non-equilibrium phase
transitions. While the equilibrium phase transitions are described on the basis of modified
Cahn-Hilliard equation, the non-equilibrium phase transitions are presented by the canonical chemical
models introduced by Schl\"ogl \cite{26}. In these models, the different ``phases'' correspond to
different stationary states of the chemical reactions system. Schl\"ogl considered two reaction
systems: the so-called ``First Schl\"ogl Reaction''\eqref{ZEqnNum382394}--\eqref{ZEqnNum219603}, which
is an analog of the second order equilibrium phase transition, and the ``Second Schl\"ogl Reaction''
\eqref{ZEqnNum655012}--\eqref{ZEqnNum594450}, which is an analog of the first order equilibrium phase
transition, for details see \cite{26}. Each of these reaction systems has four components, though the
concentrations of three reagents ( the so-called ``reservoir reagents'') are assumed to be kept constant,
and only the concentration of one reagent changes in time and space. If the system is well mixed (or there
is no spatial mass transfer), the time evolution of this reagent is governed by a nonlinear ordinary
differential equation. It is quadratic polynomial nonlinearity for the First Schl\"ogl Reaction
\eqref{ZEqnNum369199}, and the cubic nonlinearity for the Second Schl\"ogl Reaction
\eqref{ZEqnNum324453}. If the mass transfer should be taken into account, it is usually described by
diffusion equation. However, if the system is essentially inhomogeneous, e.g., undergoes a phase
transition, the proper description of the mass transfer is given by the Cahn-Hilliard equation
\cite{1,2,3,4}, complemented with nonlinear sink/source terms. For the second-order reaction system,
such an approach was pioneered by Huberman \cite{19} and Cohen and Murray \cite{20}. Apparently, being unaware
of Schl\"ogl paper, they in fact considered the interplay of equilibrium and (second-order) non-equilibrium
phase transitions. Huberman introduced Cahn-Hilliard equation with additional kinetic terms
corresponding to the reversible first-order autocatalytic chemical reaction. He analyzed the linear
stability of  stationary states and the mutual effect of spinodal decomposition and reaction. Cohen
and Murray considered the same equation in the biological context; using the nonlinear stability
analysis based on a multi-scale perturbation method, they identified bifurcations to spatial
structures. Similar equation with an additional nonlinear derivative term and an inverted sign of the
quadratic nonlinearity was used in \cite{21} to study the segregation dynamics of binary mixtures coupled
with chemical reaction. The same equation as in \cite{19,20} was used to describe the phase transitions in
chemisorbed layer \cite{22} and to model the system of cells that move, proliferate and interact via
adhesion \cite{23}. Furthermore, for the latter model, several rigorous mathematical results on the existence and
asymptotics of solutions were obtained in \cite{24,25}. On the other hand, to the best of our
knowledge, there is no study of the Cahn-Hilliard equation with the third order reaction terms in the
literature.

Our aim in the present work was to consider the possibly simple situation, where the interplay of the
equilibrium and non-equilibrium phase transitions could be observed explicitly. Thus, we considered the
advancing fronts which ``combine'', in some sense, these both transitions. We obtained several exact
travelling wave solutions, which exhibit an explicit parametric dependence. Naturally, for both
transitions to proceed simultaneously, some additional constraints should be imposed on the parameters
of the model.

To get a more direct insight here, we return to dimensional parameters. Starting from the CHHCM equation
supplemented by an additional convective term and viscosity, we see that the coexistence of equilibrium
and ``second-order'' non-equilibrium transformations in the form of a constant-velocity transition front
imposes quite rigid constraints on the parameters. From \eqref{ZEqnNum363435} the dimensional velocity
$V={vl\mathord{\left/ {\vphantom {vl \tau }} \right. \kern-\nulldelimiterspace} \tau } $ is
\begin{equation} \label{4.1} V=\sqrt{2} \frac{k'_{11} }{\sqrt{q} } \bar{\varepsilon }-\bar{\alpha }X_{1} .
\end{equation}
Here, $X_{1} =u_{1} X_{0} $ is the dimensional stationary concentration of the reaction system; from
\eqref{ZEqnNum930329} we have
\begin{equation} \label{4.2} X_{1} =u_{1} X_{0} =\frac{k_{11} A-k_{21} B}{k'_{11} } .
\end{equation}
Remarkably, in the absence of the field, $\bar{\alpha }=0$, the velocity does not depend on this
concentration, but on the parameters of the ``Cahn-Hilliard part'' $q,\varepsilon $ and on the
reaction constant for the reverse first reaction~\eqref{ZEqnNum382394} only
\begin{equation} \label{ZEqnNum984804} V=\sqrt{2} \frac{k'_{11} }{\sqrt{q} } \bar{\varepsilon }.
\end{equation}
In this case, the velocity of the anti-kink solution is always positive, while that of the kink-solution is
negative. That is, the stable state $X_{1} $ of the chemical system always spreads on the cost of the unstable
zero state. In the absence of the field and viscosity, the constraints imposed on the stationary values
of polynomial part of the chemical potential $\tilde{X}_{i} =\tilde{u}_{i} X_{0} ={\tilde{u}_{i}
\mathord{\left/ {\vphantom {\tilde{u}_{i}  \sqrt{q} }} \right. \kern-\nulldelimiterspace} \sqrt{q} } $
are very rigid indeed
\begin{equation} \label{4.4} \tilde{X}_{1} =X_{1} ;\quad \tilde{X}_{2} =\frac{1}{2} X_{1} ;\quad
\tilde{X}_{3} =0;
\end{equation}
i.e., the stable stationary states for equilibrium transition should coincide with the stationary
states for the reaction system. This also means  that the homogeneous part of the thermodynamic
potential $\Phi$ should be a symmetric function with equal-depth wells.

As already mentioned, we consider the isothermal situation only; still, it may be interesting to check
the limit of ``critical state'' for the equilibrium phase transition, i.e., for the ``Cahn-Hilliard part''.
As usually for this model, it is assumed $s\sim (T-T_c)$ [for symmetric potential this also means 
$\delta \sim (T-T_c)^{1/2}$], where $T_c$ is the critical temperature. Then, in terms of our model,
$T\rightarrow T_c$ corresponds to $\tilde{X}_{1}\rightarrow 0$. Thus, the larger equilibrium
concentration scales as
\begin{equation} \label{4.5a} \tilde{X}_{1} \sim (T-T_c)^{1/2} .
\end{equation}
From \eqref{4.2} and \eqref{4.4}, the compatibility of the transitions yields
\begin{equation} \label{4.5b} k'_{11}=\frac{k_{11} A-k_{21} B}{\tilde{X}_{1}  } .
\end{equation}
Substitution of the latter expression into \eqref{ZEqnNum984804} shows that if the equilibrium
transition approaches the critical state, the velocity of the front diverges as ${\left (\tilde{X}_{1}
\right )}^{-1}$, i.e.,
\begin{equation} \label{4.5c} V\sim (T-T_c)^{-1/2} .
\end{equation}

If the viscosity is non-zero (but still $\bar{\alpha }=0$), the expression for the velocity
\eqref{ZEqnNum984804} does not change; however, the exact expressions for the stationary values of the
polynomial part of the chemical potential become
\begin{equation} \label{4.5} \tilde{X}_{1} =X_{1} ;\, \, \, \, \tilde{X}_{2} =\frac{1}{2} X_{1} -
\frac{\bar{\eta }k'_{11} }{q} ;\, \, \, \, \, \tilde{X}_{3} =0.
\end{equation}
That is, while the  stationary states for the equilibrium transition should again coincide with the
stationary values for the reaction system, the unstable state should be shifted to the lower value. Thus,
to compensate the additional dissipation, the homogeneous part of the thermodynamic potential becomes
asymmetric, the potential well corresponding to $X_{1} $ is now deeper, see \eqref{ZEqnNum545762}; the
difference, naturally, disappears for zero viscosity $\bar{\eta }$. On the other hand, if $\,
\bar{\alpha }\ne 0;\, \, \, \bar{\eta }=0$,
\begin{equation} \label{ZEqnNum320257} \tilde{X}_{1,2} =\frac{X_{1} }{4} \left[3\pm \sqrt{1+16
\frac{\bar{\alpha }\bar{\varepsilon }}{\sqrt{2} Mq^{\frac{3}{2} } X_{1}^{2} } } \right].
\end{equation}
This means that for positive $\alpha $, the order parameter value for the final state after transition,
$X=X_{1} $, is somewhat lower than the equilibrium value $\tilde{X}_{1} $; thus, the presence of the
field can prevent final equilibration. The unstable equilibrium value $\tilde{X}_{2} $ should be
somewhat lower too, so the potential $\Phi $ is again asymmetric. Now, let both $\alpha \ne 0$, $\eta
\ne 0$. The expression for the velocity is independent of $\eta $; for the special value
\begin{equation} \label{4.7} \bar{\alpha }\, =\frac{2\sqrt{2} \bar{\varepsilon }k'_{11} }{X_{1} \sqrt{q} }
\end{equation}
the velocity is zero, i.e., for the corresponding value of the applied field, the transition front
becomes static. The latter expression depends both on the Cahn-Hilliard parameters and on $k'_{11} $
and $X_{1} $, so the static front is due to the balance of equilibrium and reactive processes. The
viscosity $\bar{\eta }$ has dropped out from the corrections to the stationary states. This is
physically reasonable: there is no dissipation for the static transition front; the deviation of the
equilibrium value $\tilde{X}_{1} $ of the order parameter from the final state after transition
$X=X_{1} $, see \eqref{ZEqnNum160622}, is exactly the same as given by \eqref{ZEqnNum320257} (i.e.,
for $\bar{\eta }=0$ case) for this special value of $\bar{\alpha }$.

In appendix we consider the model introduced by Puri and Frish \cite{21}. While it looks similar to
CHHCM equation, for their reaction system the stable state is $\psi =0$, and $\psi =1$ is unstable.
However, for the homogeneous part of the thermodynamic potential, the state $\psi =0$ is an unstable
one. Thus, for the simultaneous constant-velocity transition to exist, the stable state $\psi _{3} $
should nearly merge with zero, $\psi _{3} \simeq -0,212$, and the potential should be very far from
symmetric.

Now, considering the convective viscous CHS equation \eqref{ZEqnNum504256}, we return to dimensional
parameters again. From \eqref{ZEqnNum959690}, the dimensional velocity $V={vl\mathord{\left/ {\vphantom
{vl \tau }} \right. \kern-\nulldelimiterspace} \tau } $ is now
\begin{equation} \label{4.8} V=\sqrt{2} \frac{k'_{12} }{\sqrt{q} } \bar{\varepsilon }
\left(\frac{X_{1} }{2} -X_{2} \right)-\frac{\bar{\alpha }X_{1} }{2}
\end{equation}
As compared to the second order non-equilibrium phase transition, the situation for the first order
non-equilibrium phase transition is much more ``flexible''. As in the section~\ref{Sec:3}, we consider
first the absence of the field, $\bar{\alpha }=0$:
\begin{equation} \label{ZEqnNum319476} V=\sqrt{2} \frac{k'_{12} }{\sqrt{q} } \bar{\varepsilon }
\left(\frac{X_{1} }{2} -X_{2} \right).
\end{equation}
Comparing the latter equation with \eqref{ZEqnNum984804} we see that this expression is very similar
to the coefficient in \eqref{ZEqnNum319476} (to avoid confusion we remind that $k'_{11} $ and $k'_{12}
$ have different dimensionality). However, the dependence of velocity on the stationary values of
concentration, $X_{1} ,\, \, X_{2} ,\, \, 0,$ is exactly the same as for the well known
travelling-wave solution for the diffusion equation with cubic nonlinearity; for $X_{2} =\frac{X_{1}
}{2} $, the velocity is zero, that is the front becomes static. Moreover, for zero field, the viscosity
$\bar{\eta }$ enters the constraints \eqref{ZEqnNum719941} and \eqref{ZEqnNum552588} always multiplied
by $\left(\frac{X_{1} }{2} -X_{2} \right)$, see \eqref{ZEqnNum316998}. Particularly, for the static
front, the stationary concentrations $\tilde{X}_{1} ,\, \, \tilde{X}_{2} $ will not depend on
$\bar{\eta }$, which is reasonable physically. If additionally to $\bar{\alpha }=0$ it is also
$\bar{\eta }=0$, that is the CHS-case, the final value after transition $X_{1} $ will deviate from the
equilibrium value, see \eqref{ZEqnNum238118}. Taking into account $\bar{\varepsilon }\ll 1$, we get
\begin{equation} \label{4.10)} X_{1} \simeq \tilde{X}_{1} +2\frac{\bar{\varepsilon }^{2} k'_{12} }
{X_{1} Mq^{2} } .
\end{equation}
That is, even in the absence of the applied field and viscosity, the order parameter value for the final
state after transition, $X=X_{1} $, is somewhat higher than the equilibrium value $\tilde{X}_{1} $,
the phase is oversaturated with $X$. However, comparing the values of the $\Phi \left(X_{1} \right)$
and $\Phi \left(\tilde{X}_{1} \right)$ we see that the deviation of the thermodynamic potential from
its equilibrium value is of the higher order in $\bar{\varepsilon }$. Different from CHHCM case, for
CHS case in the limit of critical state for the equilibrium phase transition, i.e., for the
``Cahn-Hilliard part'', these transitions become incompatible. Indeed, for $T \rightarrow T_c$ we need
to take the limit $\tilde{X}_{1}\rightarrow 0$ again. However, as it is evident from
\eqref{ZEqnNum238118}, there is a non-zero lower limit for $\tilde{X}_{1}$. For smaller values, this
expression is physically senseless. Thus, if the equilibrium concentration $\tilde{X}_{1}$ of the
``Cahn-Hilliard part'' approaches this limit, the equilibrium and non-equilibrium transitions could not
proceed simultaneously, at the same front.

If $\, \bar{\alpha }\ne 0;\, \, \, \bar{\eta }=0$, see \eqref{ZEqnNum106708}, similar to convective
CHHCM, the final state after transition is slightly undersaturated by $X$ due to the presence of the
field. If both $\alpha $ and $\eta $ are non-zero, for the special value of $\bar{\alpha }$, see
\eqref{ZEqnNum936193}, the velocity is zero, i.e., for the corresponding value of the applied field, the
transition front becomes static even for $X_{2} \ne \frac{X_{1} }{2} $. Then, the viscosity $\eta $ 
self-consistently drops out from the corrections to the stationary states, see
\eqref{ZEqnNum383371}.

For the illustration purpose it is instructive to present the connection between the ``observable''
parameters, e.g., velocity $V$ of the transition front and the dimensional effective width of the
transition front $\sigma $. If $Z$ is the dimensional travelling wave coordinate, the argument of
$\tanh $ is
\begin{equation} \label{4.11}\frac{u_{1} }{2\sqrt{2} \varepsilon } z=\frac{X_{1} l}{X_{0} 2
\sqrt{2} \bar{\varepsilon }} \frac{Z}{l} =\frac{X_{1} \sqrt{q} }{2\sqrt{2} \bar{\varepsilon }} Z .
\end{equation}
Then, $\sigma =\frac{2\sqrt{2} \bar{\varepsilon }}{X_{1} \sqrt{q} } $. This expression is the same for
both reactions, though the expressions for $X_{1} $ are different, see below.

\begin{figure}[!t]
\centerline{\includegraphics[width=0.45\textwidth]{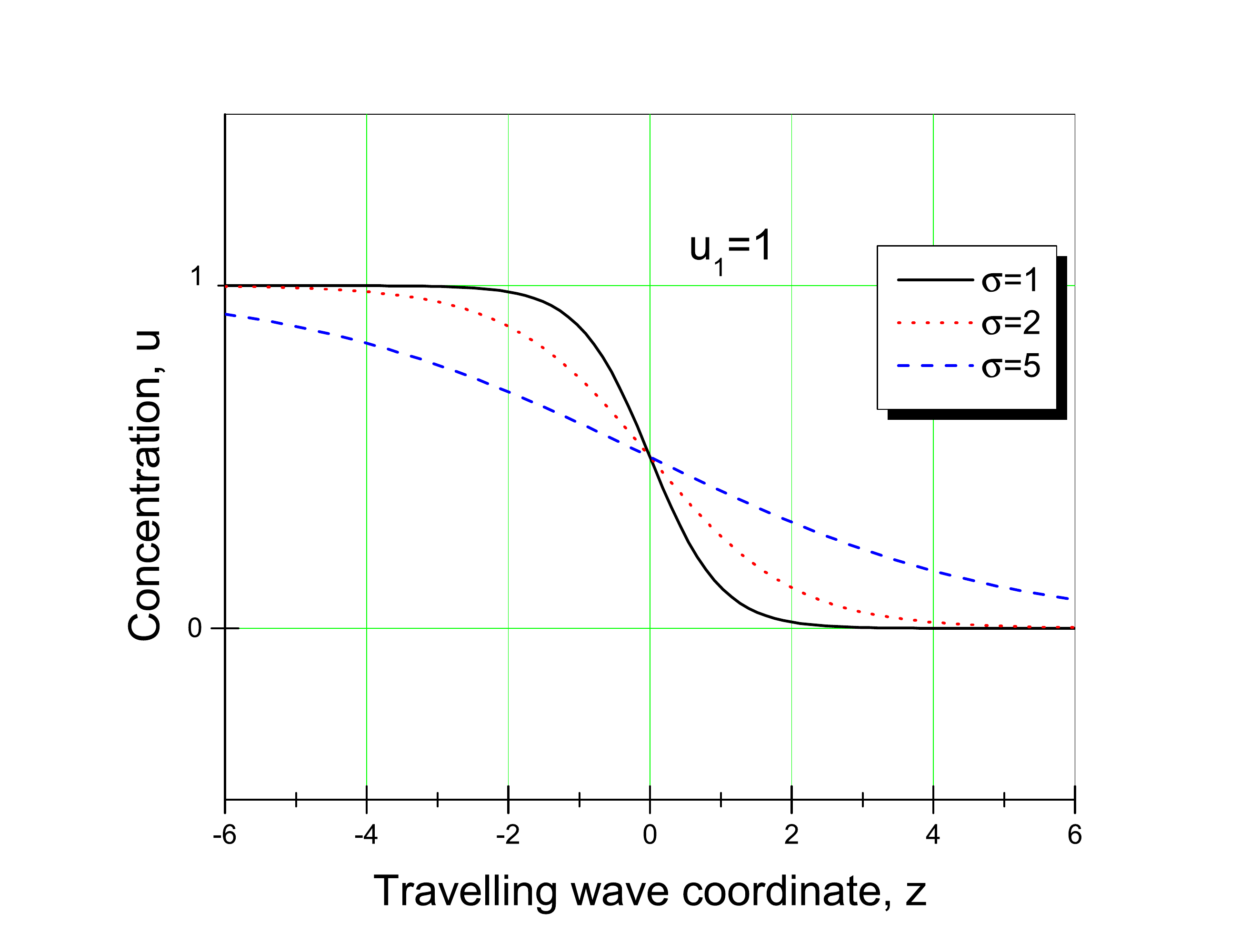}} 
\caption{(Colour online) The trevelling wave front in the
case of the 1st Schl\"ogl reaction for several wave velocities $v\sim \sigma$ and $u_1 =1$.} \label{fig.1}
\end{figure}

For the first reaction (without field)
\begin{equation}\label{4.12}
V=k'_{11} \frac{\sqrt{2} \bar{\varepsilon }}{\sqrt{q} } =k'_{11} \frac{\sigma X_{1} }{2}  .
\end{equation}
From the definition of $X_{1} $ \eqref{4.2} we find
\begin{equation}\label{4.13}
V=\frac{\sigma }{2} k_{11} A\left(1-\rho _{1} \right),
\end{equation}
where the parameter $\rho _{1} ={k_{21} B}/{k_{11} A} $ is characteristic of the first reaction:
if $\rho _{1} \to 1$, the difference between stationary states disappears, if $\rho _{1} \to 0$ this
difference is maximal, which also corresponds to the maximal velocity (figure~\ref{fig.1}).

\begin{figure}[!t]
\centerline{\includegraphics[width=0.55\textwidth]{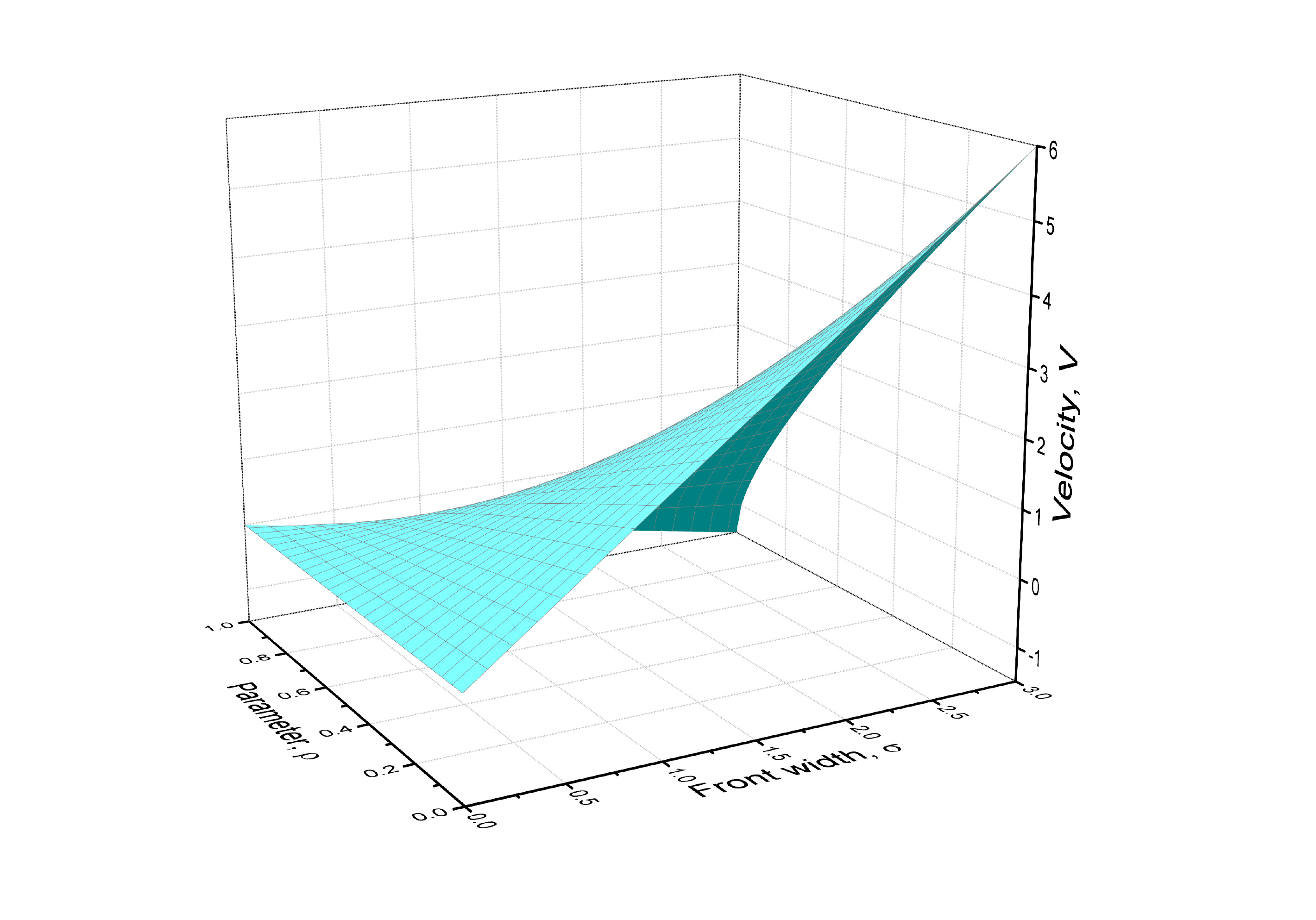}} 
\caption{(Colour online) The trevelling wave front velocity
in the case of 2nd Schl\"ogl reaction as function of the front width $\sigma $ and parameter $\rho_2$.}
\label{fig.2}
\end{figure}

For the second reaction
\begin{equation} \label{4.14}
V=\frac{\sigma k'_{12} X_{1} }{2} \left(\frac{X_{1} }{2} -X_{2} \right) \, .
\end{equation}
However, $X_{1} ,\, \, X_{2} $ are now the roots of quadratic equation:
$ {u^{2} -Ru+Q=0}$ where $R=\frac{k_{12} A}{k'_{12} X_{0} }$ and $Q=\frac{k_{22} B}{k'_{12} X_{0}^{2} }$.
Introducing a characteristic parameter for the second reaction $\rho _{2} $,
\[\rho _{2} =4\frac{k'_{12} k_{22} B}{\left(k_{12} A\right)^{2} } ;\, \, \, \, \, 0<\rho _{2} <1 ,\]
we get
\begin{equation}\label{4.15}
X_{1} =\frac{k_{12} A}{2k'_{12} } \left[1+\sqrt{1-\rho _{2} } \right];\, \, \, \, \, X_{2}
=\frac{k_{12} A}{2k'_{12} } \left[1-\sqrt{1-\rho _{2} } \right]
\end{equation}
and
\begin{equation}\label{4.16}
V=\sigma \frac{\left(k_{12} A\right)^{2} }{16k'_{12} } \left\{2\sqrt{1-\rho _{2} } -3\rho _{2} +2\right\} \, .
\end{equation}
The dependence of the wave front velocity on its width $\sigma $ and parameter $\rho_2$ according to
the above equation with $\left(k_{12} A\right)^{2} / \left(16k'_{12}\right) =1/2$ is shown in
figure~\ref{fig.2}.

Summing up, without viscosity and applied field, the constant-velocity combined-transition-front model
for CHHCM is not very instructive, both transitions should be too rigidly adjusted to each other (of
course, for the non-constant-velocity-fronts, the situation could be quite different). However, the
presence of the field and/or viscosity changes the situation; the concentration $X$ in the final state
may deviate from its equilibrium value and even the transition may be stopped. On the other hand, for
the CHS equation, the effect of the non-equilibrium transition, i.e., of the reaction system, is much
stronger. The transition front may be stopped, or even reversed both by changing the stationary states
of the reaction system and by the field. The final state may be undersaturated or oversaturated, creating
non-equilibrated phases.

\appendix

\section{Cahn-Hilliard-Puri-Frish equation} \label{Sec:A}

\renewcommand{\theequation}{A.\arabic{equation}}

Here, for completeness we give the exact travelling wave solution for one-dimensional version of
equation, introduced by Puri and Frish \cite{21}. To match our consideration with \cite{21} we use
their original notations and normalizations, which are different from these in the other parts of the
present paper. After the correction of the evident misprint and considering the generally asymmetric
thermodynamic potential, the equation (9) of \cite{21} is
\begin{equation} \label{ZEqnNum209251} \frac{\partial \psi }{\partial t} =-\frac{\partial ^{2} }
{\partial x^{2} } \left\{\psi +\delta \psi ^{2} -\psi ^{3} +\frac{\partial ^{2} \psi }{\partial x^{2} }
\right\}-\alpha \psi \left(1-\psi \right)+\frac{\alpha }{\sigma } \psi \frac{\partial ^{2}
\psi }{\partial x^{2} } .
\end{equation}
Here, $\alpha $ is a phenomenological constant, proportional to the ratio of the characteristic times for a
spin exchange and reaction, $\sigma $ is proportional to the inverse square of lattice spacing. Evidently,
dropping the last term in the right-hand side of \eqref{ZEqnNum209251} we obtain CHHCM equation,
though with an inverted sign of the reaction term, see section~\ref{Sec:2}. This means that for the
reaction system the stable state is $\psi =0$, while $\psi =1$ is unstable. Looking for the travelling
wave solutions of \eqref{ZEqnNum209251}, we introduce the travelling wave coordinate $z=x-vt$. This
yields
\begin{equation} \label{ZEqnNum650254} -v\frac{\rd\psi }{\rd z} =-\frac{\rd^{2} }{\rd z^{2} } \left\{\psi +
\delta \psi ^{2} -\psi ^{3} +\frac{\rd^{2} \psi }{\rd z^{2} } \right\}-\alpha \psi \left(1-\psi
\right)+\frac{\alpha }{\sigma } \psi \frac{\rd^{2} \psi }{\rd z^{2} } .
\end{equation}

Introducing the ansatz
\begin{equation} \label{ZEqnNum569052} \frac{\rd\psi }{\rd z} =\kappa \psi \left(1-\psi \right),
\end{equation}
which for the positive $\kappa $ and $0\leqslant \psi \leqslant 1$ corresponds to the kink-like solution, the last two
terms in the right-hand side of \eqref{ZEqnNum650254} could be rewritten as
\begin{equation} \label{5.4} -\alpha \psi \left(1-\psi \right)+\frac{\alpha }{\sigma } \psi
\frac{\rd^{2} \psi }{\rd z^{2} } =\frac{\rd}{\rd z} \left[-\frac{\alpha }{\kappa } \psi +\frac{\alpha }{2\sigma }
\kappa \psi ^{2} -\frac{2\alpha }{3\sigma } \kappa \psi ^{3} \right].
\end{equation}

Substituting the latter expression into \eqref{ZEqnNum650254}, integrating once and rearranging the
terms we get
\begin{equation} \label{5.5} \left(v-\frac{\alpha }{\kappa } \right)\psi +\frac{\alpha }{2\sigma }
\kappa \psi ^{2} -\frac{2\alpha }{3\sigma } \kappa \psi ^{3} -\frac{\rd}{\rd z} \left(\psi +\delta
\psi ^{2} -\psi ^{3} +\frac{\rd^{2} \psi }{\rd z^{2} } \right)=C_{1} .
\end{equation}
For the latter equation to be satisfied, the expression under the derivative should be a quadratic
function of $u$:
\begin{equation} \label{ZEqnNum909675}\psi +\delta \psi ^{2} -\psi ^{3} +\frac{\rd^{2} \psi }{\rd z^{2} } =
\beta \psi +\gamma \psi ^{2} +C_{2} \, ,
\end{equation}
\begin{equation} \label{ZEqnNum861166} \left(v-\frac{\alpha }{\kappa } \right)\psi +\frac{\alpha }
{2\sigma } \kappa \psi ^{2} -\frac{2\alpha }{3\sigma } \kappa \psi ^{3} -\left(\beta +2\gamma
\psi \right)\frac{\rd\psi }{\rd z} =C_{1}  ,
\end{equation}
where $\beta ,\, \, \gamma ,\, \, C_{1} $ and $C_{2} $ are constants. Substitution of the
corresponding expressions for the derivatives into \eqref{ZEqnNum909675} and \eqref{ZEqnNum861166}
yields
\begin{equation} \label{5.8} \left(v-\frac{\alpha }{\kappa } \right)\psi +\frac{\alpha }{2\sigma }
\kappa \psi ^{2} -\frac{2\alpha }{3\sigma } \kappa \psi ^{3} +\kappa \left[2\gamma \psi ^{3} +
\left(\beta -2\gamma \right)\psi ^{2} -\beta \psi \right]=C_{1} ,
\end{equation}
\begin{equation} \label{5.9} \psi +\delta \psi ^{2} -\psi ^{3} +\kappa ^{2} \left(2\psi ^{3} -3
\psi ^{2} +\psi \right)=\beta \psi +\gamma \psi ^{2} +C_{2} .
\end{equation}

Rearranging and equating to zero coefficients at all powers of $\psi $, we obtain the system of
constraints imposed on the parameters.
\begin{equation} \label{ZEqnNum286581} 2\kappa ^{2} =1,
\end{equation}
\begin{equation} \label{5.1)} \delta -\frac{3}{2} =\gamma ,
\end{equation}
\begin{equation} \label{ZEqnNum359737} \beta =\frac{3}{2} ,
\end{equation}
\begin{equation} \label{5.13} \gamma =\frac{\alpha }{3\sigma } ,
\end{equation}
\begin{equation} \label{5.14} \frac{\alpha }{2\sigma } +\beta -2\gamma =0,
\end{equation}
\begin{equation} \label{ZEqnNum245489} v=\frac{\alpha }{\kappa } +\kappa \beta .
\end{equation}
If these constraints are satisfied, the solutions of \eqref{ZEqnNum569052} are simultaneously the
solutions of the travelling-wave equation \eqref{ZEqnNum650254}, which, quite analogously to
\eqref{ZEqnNum151330} is
\begin{equation} \label{5.16} \psi =\frac{1}{2} \left\lbrace 1+\tanh\left[\frac{1}{2\sqrt{2} }
\left(x-vt\right)\right]\right\rbrace .
\end{equation}

From \eqref{ZEqnNum286581}, \eqref{ZEqnNum359737} and \eqref{ZEqnNum245489} $v=\sqrt{2} \alpha
+\frac{3}{2\sqrt{2} } $, the velocity of the kink is always positive ($\alpha $ is per definition
positive), i.e., the stable $\psi =0$ state spreads on the cost of the unstable state $\psi =1$. The
constraints imposed on the parameters of the model are $\alpha =9\sigma ;\, \, \, \delta
={9\mathord{\left/ {\vphantom {9 2}} \right. \kern-\nulldelimiterspace} 2} $. Then, the roots of
equation
\begin{equation} \label{5.17} \psi ^{3} -\delta \psi ^{2} -\psi =0
\end{equation}
are $\psi _{1} \simeq 4,712;\, \, \psi _{2} \, =0;\, \, \, \psi _{3} \simeq -0,212$, i.e., the
potential $\Phi \left(\psi \right)$ should be very far from symmetric.
\newline

\section*{Acknowledgements}
We are thankful to O.S. Bakai for the attention to the paper and valuable remarks.

{\small \topsep 0.6ex

}

\ukrainianpart

\title{Модель Кана-Гіліярда з реакціями Шльогля: взаємодія рівноважного і нерівноважного
фазових переходів. \\І. Розв'язок типу рухомої хвилі}
\author{П.О. Мчедлов-Петросян, Л.М. Давидов}
\address{Національний науковий центр ``Харківський фізико-технічний інститут'', Харків, 61108,
вул. Академічна, 1
}
%
%
%

\makeukrtitle 

\begin{abstract}
\tolerance=3000%
Робота присвячена моделюванню взаємодії рівноважних і нерівноважних фазових переходів. Рівноважні
фазові переходи моделюються модифікованим рівнянням Кана-Хільярда, а нерівноважні --- системою
хімічних реакцій Шльогля. Ми розглядаємо фронти, що поширюються, які поєднують обидва ці фазові
переходи. Отримано розв'язки типу рухомої хвилі. Умови їхнього існування і залежності від параметрів
моделей аналізуються детально. Обговорюється можливість існування нерівноважних фаз.
\keywords фазовий перехід, нерівноважний фазовий перехід, рівняння Кана-Хільярда, реакції
Шльогля
\end{abstract}

\end{document}